\RequirePackage[columnwise]{lineno}
\documentclass[aps,prl,twocolumn,superscriptaddress,footinbib,linenumbers]{revtex4}  
\usepackage{graphicx}  
\usepackage{dcolumn}   
\usepackage{bm}        
\usepackage{braket}
\usepackage{exscale}                  
\usepackage[intlimits]{amsmath}       
\usepackage{amsfonts}
\usepackage{amssymb,amscd}
\usepackage{color}
\usepackage{hyperref}
\usepackage{subfigure}
\usepackage[normalem]{ulem}
\usepackage{slashed}
\usepackage{leftidx}
\usepackage{xcolor,soul}
\soulregister\cite7
\soulregister\ref7

\begin{document}

\title{Realization of a Bosonic Antiferromagnet}

\author{Hui~Sun}
\affiliation{Hefei National Laboratory for Physical Sciences at Microscale and Department of Modern Physics, University of Science and Technology of China, Hefei, Anhui 230026, China}
\affiliation{Physikalisches Institut, Ruprecht-Karls-Universit\"{a}t Heidelberg, Im Neuenheimer Feld 226, 69120 Heidelberg, Germany}
\affiliation{CAS Centre for Excellence and Synergetic Innovation Centre in Quantum Information and Quantum Physics, University of Science and Technology of China, Hefei, Anhui 230026, China}

\author{Bing~Yang}
\email[e-mail:]{bing@physi.uni-heidelberg.de}
\affiliation{Hefei National Laboratory for Physical Sciences at Microscale and Department of Modern Physics, University of Science and Technology of China, Hefei, Anhui 230026, China}
\affiliation{Physikalisches Institut, Ruprecht-Karls-Universit\"{a}t Heidelberg, Im Neuenheimer Feld 226, 69120 Heidelberg, Germany}
\affiliation{Institut f\"ur Experimentalphysik, Universit\"at Innsbruck, Technikerstra{\ss}e 25, 6020 Innsbruck, Austria}

\author{Han-Yi~Wang}
\affiliation{Hefei National Laboratory for Physical Sciences at Microscale and Department of Modern Physics, University of Science and Technology of China, Hefei, Anhui 230026, China}
\affiliation{Physikalisches Institut, Ruprecht-Karls-Universit\"{a}t Heidelberg, Im Neuenheimer Feld 226, 69120 Heidelberg, Germany}
\affiliation{CAS Centre for Excellence and Synergetic Innovation Centre in Quantum Information and Quantum Physics, University of Science and Technology of China, Hefei, Anhui 230026, China}

\author{Zhao-Yu~Zhou}
\affiliation{Hefei National Laboratory for Physical Sciences at Microscale and Department of Modern Physics, University of Science and Technology of China, Hefei, Anhui 230026, China}
\affiliation{Physikalisches Institut, Ruprecht-Karls-Universit\"{a}t Heidelberg, Im Neuenheimer Feld 226, 69120 Heidelberg, Germany}
\affiliation{CAS Centre for Excellence and Synergetic Innovation Centre in Quantum Information and Quantum Physics, University of Science and Technology of China, Hefei, Anhui 230026, China}

\author{Guo-Xian~Su}
\affiliation{Hefei National Laboratory for Physical Sciences at Microscale and Department of Modern Physics, University of Science and Technology of China, Hefei, Anhui 230026, China}
\affiliation{Physikalisches Institut, Ruprecht-Karls-Universit\"{a}t Heidelberg, Im Neuenheimer Feld 226, 69120 Heidelberg, Germany}
\affiliation{CAS Centre for Excellence and Synergetic Innovation Centre in Quantum Information and Quantum Physics, University of Science and Technology of China, Hefei, Anhui 230026, China}

\author{Han-Ning~Dai}
\affiliation{Hefei National Laboratory for Physical Sciences at Microscale and Department of Modern Physics, University of Science and Technology of China, Hefei, Anhui 230026, China}
\affiliation{CAS Centre for Excellence and Synergetic Innovation Centre in Quantum Information and Quantum Physics, University of Science and Technology of China, Hefei, Anhui 230026, China}

\author{Zhen-Sheng~Yuan}
\email[e-mail:]{yuanzs@ustc.edu.cn}
\affiliation{Hefei National Laboratory for Physical Sciences at Microscale and Department of Modern Physics, University of Science and Technology of China, Hefei, Anhui 230026, China}
\affiliation{Physikalisches Institut, Ruprecht-Karls-Universit\"{a}t Heidelberg, Im Neuenheimer Feld 226, 69120 Heidelberg, Germany}
\affiliation{CAS Centre for Excellence and Synergetic Innovation Centre in Quantum Information and Quantum Physics, University of Science and Technology of China, Hefei, Anhui 230026, China}

\author{Jian-Wei~Pan}
\email[e-mail:]{pan@ustc.edu.cn}
\affiliation{Hefei National Laboratory for Physical Sciences at Microscale and Department of Modern Physics, University of Science and Technology of China, Hefei, Anhui 230026, China}
\affiliation{Physikalisches Institut, Ruprecht-Karls-Universit\"{a}t Heidelberg, Im Neuenheimer Feld 226, 69120 Heidelberg, Germany}
\affiliation{CAS Centre for Excellence and Synergetic Innovation Centre in Quantum Information and Quantum Physics, University of Science and Technology of China, Hefei, Anhui 230026, China}


\maketitle

\textbf{
Quantum antiferromagnets are of broad interest in condensed matter physics as they provide a platform for studying exotic many-body states \cite{Auerbach:2012} including spin liquids \cite{Savary:2016} and high-temperature superconductors \cite{Lee:2006}.
Here, we report on the creation of a one-dimensional Heisenberg antiferromagnet with ultracold bosons.
In a two-component Bose-Hubbard system, we switch the sign of the spin-exchange interaction and realize the isotropic antiferromagnetic Heisenberg model in an extended 70-site chain.
Starting from a low-entropy N\'eel-ordered state, we use optimized adiabatic passage to approach the bosonic antiferromagnet.
We demonstrate the establishment of antiferromagnetism by probing the evolution of the staggered magnetization and spin correlations of the system.
Compared with condensed matter systems, ultracold gases in optical lattices can be microscopically engineered and measured, offering significant advantages for exploring bosonic magnetism and spin dynamics \cite{Gross:2017}.
}


For the origin of quantum magnetism in cold-atom simulator \cite{Gross:2017}, the fundamental symmetry of the constituent particles supports the general form of the spin-exchange interactions, ferromagnetic for bosons and antiferromagnetic for fermions \cite{Duan:2003}.
Interestingly, the sign of the exchange interactions \cite{Trotzky:2008} and the motional degrees of freedom of atoms can be engineered by precisely controlling a staggered lattice potential, resulting in an antiferromagnetic spin model for bosons.
In recent years, antiferromagnetic spin correlations have been experimentally observed in the fermionic quantum gases \cite{Greif:2013,Murmann:2015,Hart:2015,Boll:2016,Cheuk:2016,Mazurenko:2017,Drewes:2017}.
However, the bosonic antiferromagnet has not been realized.
Compared with fermionic atoms, besides the advantage of readily achieved lower entropy in bosonic quantum gases \cite{Yang:2020}, the tunable intra- or inter-spins interactions lead to rich phase diagrams \cite{Duan:2003,Altman:2003}.
Additionally, unlike in Fermi-Hubbard systems where the sign problem prevents their numerical solutions, our bosonic models can enable benchmarks of numerical methods at the current state of quantum simulation in noisy intermediate-scale devices \cite{Hauke:2012}.
Making use of a defect-free Mott insulator \cite{Yang:2020}, one promising approach to achieve the many-body state is via the adiabatic transformation \cite{Sorensen:2010,Chiu:2018}.
Experimental challenges lie in engineering the Hamiltonian of two-component bosons in a many-body system and maintaining adiabaticity during the state preparation.


\begin{figure}[!t]
\centering
\includegraphics[width=80mm]{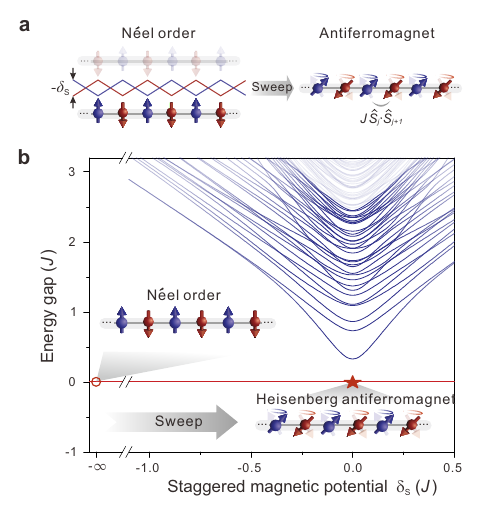}
\caption{Adiabatic passage for creating a bosonic antiferromagnet.
(a) Adiabatic protocol. The antiferromagnet is approached by slowly turning on the Heisenberg interaction and simultaneously sweeping a staggered magnetic potential.
(b) Ground-state passage. Energy levels of the antiferromagnetic Heisenberg model are shown in the unit of superexchange coupling $J$, exemplified for a 10-site spin chain.
The sweeping starts from a N\'eel-order (circle $\delta_s \rightarrow -\infty$) and ends up with the highly-entangled Heisenberg antiferromagnet with the SU(2) symmetry (star $\delta_s = 0$).
}\label{fig1}
\end{figure}


Here, we adiabatically prepare the one-dimensional (1D) antiferromagnet following the lowest-energy passage of a Heisenberg spin Hamiltonian.
The experiment begins with a two-dimensional (2D) Mott insulator of $^{87}$Rb in the hyperfine state $5S_{1/2}\left|F=1, m_F=-1 \right>$  with an average filling factor of 0.992(1), which is realized based on a recently developed cooling method \cite{Yang:2020}.
The hyperfine states $\left|F=1, m_F=-1 \right>$ and $\left|F=2,m_F=-2\right>$ are defined as pseudo-spin components $\ket{\downarrow}$ and $\ket{\uparrow}$, respectively.
We create a spin chain in N\'eel order by alternatively addressing these pseudo-spin states.
On this basis, Figure \ref{fig1} shows that the target state can be approached by sweeping an effective staggered field \cite{Yang:2017a}.
We benchmark the implementation of an isotropic Heisenberg spin model by measuring its spin-relaxation dynamics.
The antiferromagnet is prepared by slowly turning on the spin interactions and gradually reducing the staggering amplitude $|\delta_s|$ until the staggered magnetization ($M_z$) vanishes.
We characterize the antiferromagnet by measuring local spin correlations and global spin fluctuations, reflecting a short-range antiferromagnetism in this 1D spin chain.
Furthermore, we investigate its spin-rotational symmetry and robustness against dissipation.

\begin{figure}
\centering
\includegraphics[width=80mm]{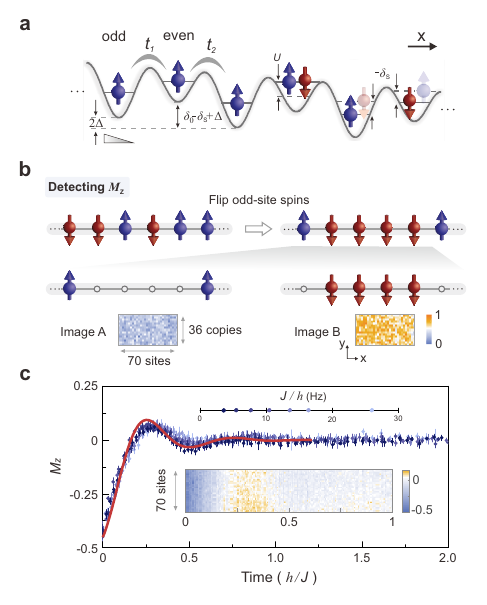}
\caption{Spin-relaxation dynamics of the many-body Heisenberg model.
(a) In a two-component Bose-Hubbard chain, we sketch all the tunable parameters of the Hamiltonian~(\ref{eq:BH}).
(b) Detection of $M_z$.
By flipping the odd-site spins, we record $M_z$ on two successive images, where the atomic densities in the ROI are selected for analysis.
(c) Relaxation of the N\'eel-ordered state.
We use various blue colors to represent the spin dynamics at different coupling strength $J$.
The solid curve shows the theoretical prediction which reaches $1.2 h/J$ with our computational capability.
In the inset, we show the spatial-resolved $M_z$ averaged over the 36 copies.
Error bars are the standard deviations throughout this Letter.
}\label{fig2}
\end{figure}

The two-component bosons in our state-dependent optical superlattice can be well described by the Bose-Hubbard Hamiltonian,

\begin{equation}
\label{eq:BH}
\begin{aligned}
&\hat{H}_{\mathrm{BH}} = -\sum_{i,\sigma}\left(t_1\hat{a}_{2i-1,\sigma}^{\dagger} \hat{a}_{2i,\sigma} +t_2\hat{a}_{2i,\sigma}^{\dagger} \hat{a}_{2i+1,\sigma}+ \mathrm{H.c.}\right) + \\
 & \sum_{j,\sigma} \left\{\frac{U}{2}\hat{n}_{j,\sigma} (\hat{n}_{j,\sigma}-1)+\left[\frac{(-1)^j}{2}\left( \delta_0 -\delta_s \sigma \right)-j\Delta\right]\hat{n}_{j,\sigma}\right\}.
\end{aligned}
\end{equation}
Here, the spins $\{\ket{\uparrow}, \ket{\downarrow}\}$ are indexed by $\sigma=\{1,-1\}$;
$\hat{a}_{j,\sigma}$ ($\hat{a}_{j,\sigma}^{\dagger} $) is the annihilation (creation) operator of an atom on site $j$, and $\hat{n}_{j,\sigma} = \hat{a}_{j,\sigma}^{\dagger} \hat{a}_{j,\sigma}$.
In superlattice, $t_1$ ($t_2$) corresponds to the tunneling strengths on odd-even (even-odd) links;
$U$ represents the on-site interaction, whose spin-dependent effect is insignificant.
Figure \ref{fig2}a illustrates the staggered energy biases between adjacent sites, including a spin-independent term $\delta_0$ and a spin-dependent term $\delta_s$.
Additionally, a weak linear potential is applied to the 1D chain, with $\Delta \ll \delta_0$.

In a normal optical lattice, where $\delta_0=0$ and $t_1=t_2=t$, the general Hubbard model can be mapped onto a Heisenberg-type spin model \cite{Duan:2003}.
The antiferromagnetic interaction (coupling strength $4t^2/U$) is assigned for fermions, while bosons have the ferromagnetic coupling $-4t^2/U$.
However, in the staggered spin-independent potential, the superexchange couplings for bosons become $J_1=-4t_1^2U/(U^2-\delta_0^2)$ $\left[ \text{or}\ J_2=-4t_2^2U/(U^2-\delta_0^2)\right]$ \cite{Trotzky:2008}.
To this end, we set $U < \delta_0$ to reverse the sign of the superexchange coupling for both types of links, realizing the antiferromagnetic interaction with bosons throughout the chain.
Meanwhile, this staggered potential localizes static motional impurities (sites filled by 0 or 2 atoms) and turns the bosonic $t-J$ model \cite{Lee:2006} into a spin model.
Besides, we control the linear potential to unify the inter-well superexchange couplings for those two links.

Therefore, in the regime of $t_1,t_2 \ll U < \ \delta_0$, our system is described by the antiferromagnetic Heisenberg model after applying the Schrieffer-Wolff transformation \cite{Auerbach:2012}.

\begin{equation}
\label{eq:Spin}
\hat{H}=\sum_{j}\left[J\hat{S}_{j}\cdot\hat{S}_{j+1} + \delta_s (-1)^{j+1}\hat{S}_j^z \right],
\end{equation}
where $\{\hat{S}_j, \hat{S}_j^z \}$ are spin-$1/2$ operators, and $J=\sqrt{J_1J_2}$ represents the superexchange coupling strength by neglecting the difference between $J_1$ and $J_2$.
The staggered magnetic field originates from the spin-dependent term in the Hamiltonian~(\ref{eq:BH}).
For characterizing this $N$-site antiferromagtism, the staggered magnetization is introduced as, $M_z=\langle\hat{M}_z\rangle=1/N\sum_{j}(-1)^j\langle\hat{S}_j^z\rangle$.


When $\delta_s \rightarrow -\infty$, the ground state of $\hat{H}$ is a N\'eel-ordered state as $\ket{\uparrow\downarrow\uparrow\downarrow\cdots}$, see Fig.~\ref{fig1}.
In this spin-balanced system, the lowest-energy excitation corresponds to flipping an arbitrary spin pair, like $\ket{\downarrow\uparrow\uparrow\downarrow\cdots}$.
The opposite N\'eel order state, $\ket{\downarrow\uparrow\downarrow\uparrow\cdots}$, represents the ground state at $\delta_s \rightarrow +\infty$.
The ground state at $\delta_s = 0$ is our target antiferromagnet $\ket{\psi_{AF}}$.
This is a highly-entangled many-body state, exhibiting zero $M_z$ but strong spin correlations.
Due to the SU(2) symmetry of $\hat{H}$, this ground state naturally gains the spin-rotational invariance.
The ground and the first-excited states are separated by a minimal energy gap of $\Delta_{ge} \propto J/N$.
The state $\ket{\psi_{AF}}$ can be approached by slowly sweeping the $\delta_s$ from $-\infty$ to 0, while the small energy gap and the intrinsic heating \cite{Ho:2007,Yang:2020} impose limitations on the adiabaticity.


\begin{figure*}
\centering
\includegraphics[width=145mm]{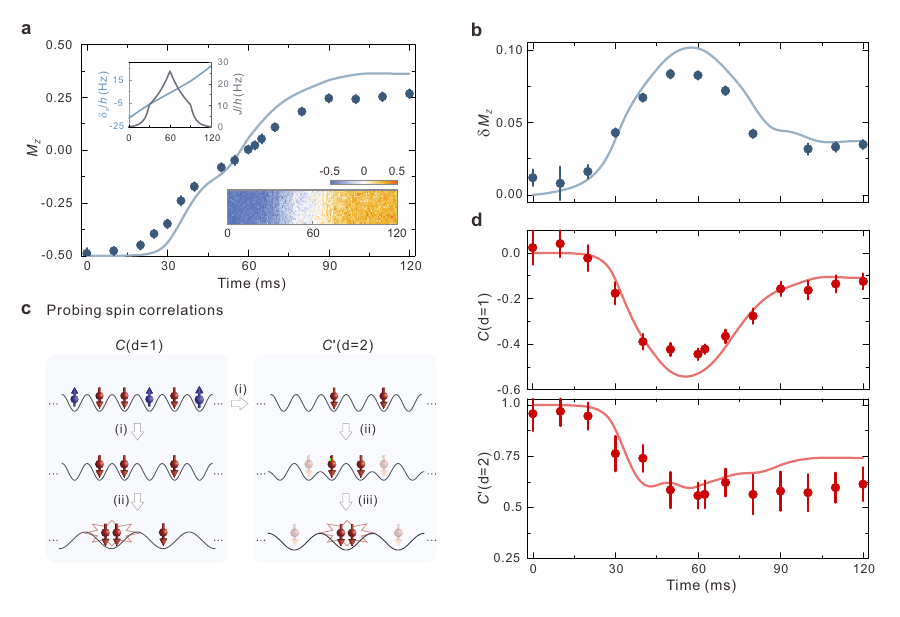}
\caption{Preparation and detection of the antiferromagnet.
(a) Evolution of $M_z$. The ramping curves of $\delta_s$ and $J$ are plotted in the upper-left inset. The lower-right inset shows the spatial-resolved $M_z$ throughout the sweep.
(b) luctuations of $M_z$.
Taking account of the detection noise and spatial constrains, we obtain $\delta M_z$ in the 70-site chains.
The blue curves in (a) and (b) are the theoretical predictions by calculating the spin model.
(c) Probing the nearest- and next-nearest-neighbor spin correlations.
The spin correlations $C(d=1)$ and $C'(d=2)$ are probed by engineering the atomic states followed with a detection of doublons (see text for details).
(d) Evolution of spin correlations during the adiabatic passage.
The solid curves are the numerical calculations of our 70-site spin chain, according to the ramp curve in (a).
C(1) equals 0 in the absence of spin correlations.
}
\label{fig3}
\end{figure*}


In the experiment, atoms are confined by a superlattice potential along the $x$-axis, as $V(x)= V_s\cos^2(2\pi x/\lambda_s) - V_l\cos^2(\pi x/\lambda_s +\theta)$.
Here, the wavelengths of short- and long-lattice lasers are $\lambda_s=$767 nm and 1534 nm;
$V_{s,l}$ are lattice depths and $\theta$ represents their relative phase.
A deep short-lattice along the $y$-axis isolates the system into copies of 1D chains.
To initialize the N\'eel order, a spin-dependent superlattice is applied to split the degeneracy of the hyperfine-transition frequencies between odd and even sites \cite{Yang:2017a}.
Then all the atoms resided on odd sites are flipped to $\ket{\uparrow}$.
Here, a region of interest (ROI) containing $70 \times 36$ sites is selected for analysis.
Consequently, 36 copies of 70-site 1D chains are prepared in the N\'eel order $\ket{\uparrow\downarrow\uparrow\downarrow\cdots}$ with an entropy per particle of $\mathcal{S}/N = 0.08(1) k_B$ ($k_B$ is the Boltzmann constant), acting as the ground state of the spin model at $\delta_s \rightarrow -\infty$.


\begin{figure}
\centering
\includegraphics[width=78mm]{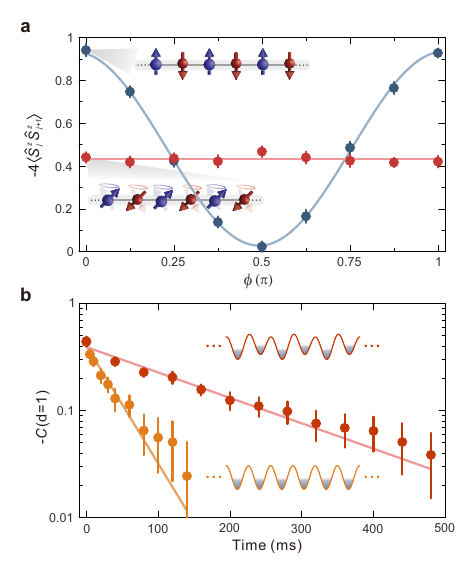}
\caption{Rotational symmetry and robustness of the state.
(a) After applying spin rotations, we measure the nearest-neighbor spin correlation of the created antiferromagnet at 60 ms in the adiabatic passage(red) and the initial N\'eel-ordered state (blue). The solid curves are sinusoidal fittings on the experimental results.
(b) The red and orange circles describe the decoherence of the antiferromagnet in a staggered lattice and a normal lattice. Their spin-exchange strengths are 26.0(5) Hz and 27.3(5) Hz, respectively. The solid curves are exponential-decay fittings of the corresponding measurements.}\label{fig4}
\end{figure}


We engineer the spin-dependent superlattice potential to implement the many-body spin model.
To enable the inter-well coupling \cite{Trotzky:2008,Greif:2013,Yang:2020}, the most pronouncing staggered potential emerges at $\theta = \pi/4$.
However, such a superlattice gives rise to insignificant spin-dependent effects.
To ensure sufficient values of both $\delta_0$ and $\delta_s$, we set the phase to $\theta=\pi/3$ and turn on the spin-dependent effect.
For compensating the discrepancy between the superexchange couplings in two types of links, we employ the gravity to project a linear potential on the 1D chain.
Furthermore, the Hubbard parameters $t_1$, $t_2$ and $U$ are mainly controlled by tuning the depths of lattices.

We investigate the spin relaxations by quenching the system to activate the Heisenberg model at $\delta_s =0$.
The $M_z$ is measured by utilizing the same sub-wavelength addressing technique, whereafter the atomic densities $n_{A, B}$ of the spin components are recorded on two images A and B, see Figure \ref{fig2}b.
Thereby, the local averaged staggered magnetization is expressed as, $M_z = (n_A-n_B)/\left[2(n_A+n_B)\right]$.
Figure \ref{fig2}c shows the spin dynamics with $J$ ranging from 3.6(1) Hz to 26.0(5) Hz, which collapse into a single curve when the evolution time is rescaled by $h/J$.
Rather than a barely incoherent decay, a coherent oscillatory spin evolution is clearly observed in this many-body system.
We perform the time-adaptive density matrix renormalization group ($t$-DMRG) method to calculate the dynamics of this 70-site spin chain.
Our results agree excellently with the theoretical predictions, indicating a faithful realization of the many-body antiferromagnetic model.

We realize the antiferromagnet by ramping $\delta_s$ and $J$ simultaneously, whose ratio $\delta_s/J$ determines the adiabaticity. One notable property of the antiferromagnet is the vanished $M_z$. However, a high-entropy state which remarkably populates the high-energy excited levels could also manifest $M_z=0$. To optimize the adiabatic passage, we continue the ramping and reach a large positive $\delta_s/J$. From the revival of $M_z$, we could infer the entropy of the target system, which simply increases as more high-energy levels are populated. Figure \ref{fig3}a shows the optimized ramping process, which begins with $\delta_s/J = -73(2)$ and the system finally achieves a low-entropy antiferromagnet within 60 ms.

We observe an enhancement of spin fluctuations during the establishment of the antiferromagnetic state. The fluctuations of $M_z$ can be formulated as $\delta M_z = \sqrt{\langle\hat{M}_z^2\rangle-\langle\hat{M}_z\rangle^2}$. We extract $\delta M_z$ by subtracting the detection noise and spatial broadening effects from the measurements (see methods). Figure \ref{fig3}b shows that $\delta M_z$ grows up gradually during the 60 ms sweep. The fluctuation decays as the ramp process continues towards the positive $\delta_s/J$.

The state is quantified by measuring its nearest- and next-nearest-neighbor spin correlations.
The correlation function is expressed as, $C(d)=4/N\sum_j \langle\hat{S}_j^z\hat{S}_{j+d}^z\rangle-\langle\hat{S}_j^z\rangle\langle\hat{S}_{j+d}^z\rangle$, where $d$ is the distance between lattice sites.
Here, we probe the spin correlations by first manipulating the atomic states in the superlattice, and then detecting double occupations with a photoassociation process \cite{Yang:2020}.
The procedure for extracting $C(1)$ is sketched as (i)-(ii) in Figure \ref{fig3}c. (i) The spin component $\ket{\uparrow}$ is removed from the lattice confinement. (ii) We combine the atoms on the neighboring sites with the superlattice.
The doublons are then removed by a photoassociation laser.
Therefore, the probability of the spin states $\ket{\downarrow\downarrow}$ equals to the atom-loss in this stage.
The other spin states can be detected by rotating them to $\ket{\downarrow\downarrow}$ before the step (i).
Thus, $C(1) = P_{\ket{\uparrow\uparrow}} + P_{\ket{\downarrow\downarrow}} - P_{\ket{\uparrow\downarrow}} - P_{\ket{\downarrow\uparrow}} + 4M_z^2$.
Figure \ref{fig3}d shows the establishment of this spin correlation, which reaches the largest value $C(1)=-0.44(3)$ at 60 ms.

Similarly, Figure \ref{fig3}c shows the procedure for probing the next-nearest-neighbor correlation.
In (i), we remove the atoms resided on the even sites and all the spin-$\ket{\uparrow}$ atoms on the odd sites.
(ii) The atoms are split in one superlattice configuration with $\theta=0$, where they have the same probability to enter each side.
(iii) We combine the atoms with the double-well superlattice at $\theta=\pi/2$. The atom correlation is also measured by counting the photoassociation-induced particle loss.
Actually, this probing method can only be used to detect the correlation between identical spins in the next-nearest-neighbor sites, defined as $C'(2) = P_{\ket{\uparrow0\uparrow}} + P_{\ket{\downarrow0\downarrow}} $.
Figure \ref{fig3}d shows $C'(2)$ during the spin dynamics, where the state probabilities are corrected by incorporating the particle-hole excitations.
At 60 ms, the correlation $C'(2) = 0.56(6)$ shows that the spin orientations tend to be identical in the next-nearest-neighbor sites.
Here, as the staggered magnetization vanishes, the spin correlations indicate the existence of entanglement among the spin chain.

Ideally, a fully adiabatic sweep will end up with the opposite N\'eel order state, like $\ket{\downarrow\uparrow\downarrow\uparrow\cdots}$.
However, in Figure \ref{fig3}, our numerical calculations show imperfect revival of the state, which is due to the nonadiabaticity of the sweep.
Besides, our experiment has some discrepancy with the numerics, which can be attributed to the motional excitations in the Bose-Hubbard chain as well as the intrinsic heating arising from the incoherent laser scattering \cite{Yang:2020}.
Comparing the measured $C(1)$ with a finite-temperature quantum Monte Carlo calculation of the spin chain, we can infer an effective entropy per particle of $\mathcal{S}/N=0.42(5) k_B$ (see methods) and a temperature of $k_B T= 0.53(7) J$.
The temperature is still far larger than the energy gap around the critical point.
However, the zero-temperature simulation is in good agreement with the experiment measurement, suggesting that such a finite temperature does not affect the adiabatic passage.
Due to the enhanced quantum fluctuations in 1D systems, the correlation function follows a power-law decay behavior. Our theoretical simulation shows that the spin correlation is short-range and the correlation length does not diverge even when the temperature approaches zero (see method).

As the antiferromagnet is generated at 60 ms of the adiabatic passage, we observe its spin rotational symmetry by measuring the spin correlation after global spin rotations.
When a Rabi pulse is applied on this state, all spins are rotated around $x$-axis by an angle $\phi \in [0, \pi]$.
Subsequently, the spin correlation $-4\langle\hat{S}^z_j\hat{S}^z_{j+1}\rangle$ is extracted.
Figure \ref{fig4}a shows that this correlation function maintains constant for the antiferromagnet.
For comparison, the initial N\'eel-ordered state is also measured following the same procedure.
In the absence of entanglement between atoms, it behaves like a single spin under rotations.
The spin-rotational invariance of our state is consistent with the SU(2) symmetry of the Hamiltonian.

The robustness of the antiferromagnet is measured by holding the atoms in the strong-coupling regime, where $J/h=26.0(5)$ Hz and $\delta_s=0$.
Figure \ref{fig4}b shows the slow decay of $C(1)$, whose lifetime 183(13) ms is 4.8(4) times of the exchange period $h/J$.
The staggered potential freezes the motions of the particles, and thus the decay is mainly caused by spin excitations \cite{Dimitrova:2020}.
This further explains the strong correlations of the state prepared through the adiabatic sweeping.
For comparison, we also monitor the damping of the correlation under the ferromagnetic spin-exchange coupling in the single-frequency lattice.
As shown in Figure \ref{fig4}b, the nearest-neighbor correlation decays much faster and the lifetime 39(4) ms equals to 1.1(1)$h/J$ and 3.0(3)$h/t$.
In the absence of staggered potential as $\delta_0=0$, the dynamics of the system is governed by the bosonic $t-J$ model \cite{Lee:2006}, where the free propagation of the particle-hole excitations makes the antiferromagnet melt rapidly.
Therefore, our staggered potential represents a powerful tool for preserving the spin correlations, which could be used for protecting the spin states in other types of Hubbard systems or the highest-energy state of spin models \cite{Sorensen:2010}.
In a doped antiferromagnet, by controlling this staggered potential, we can transform the Heisenberg spin model to a $t-J$ model and activate the interplay between spin and motional degrees of freedom for investigating the out-of-equilibrium spin dynamics.

In summary, we have demonstrated the creation of a Heisenberg antiferromagnet with ultracold bosons in optical lattices.
Our adiabatic passage circumvents the difficulties of directly cooling the atoms in the gapless phases.
In the near future, interesting topics include studying the competition between spin and charge excitations during the build-up of the antiferromagnet \cite{Hilker:2017,Chiu:2019}, cooling the spin degrees of freedom \cite{Yang:2020} and detecting the thermalization of the antiferromagnetic spin chain \cite{Pozsgay:2014}.
Based on the low entropy achieved in such a large system, we can further investigate the transport properties of the spin or charge excitations in the long-wavelength limit \cite{Haldane:1981,Giamarchi:2004,Nichols:2019,Brown:2019,Vijayan:2020,Jepsen:2020}.
Besides, the multipartite entanglement in the antiferromagnet could be certified aiming for applications in quantum metrology \cite{Almeida:2020}.
Furthermore, by subtly controlling the spin gaps \cite{White:1994} and tailoring our uniform system with a recently developed addressing technique \cite{Yang:2020s}, our method could be extended to higher-dimensional systems, to explore the resonating valence-bond states \cite{Nascimbene:2012} and also topological phases stabilized by the multi-body interactions \cite{Dai:2017}.

\begin{acknowledgments}
\textbf{Acknowledgement}
We thank Philipp Hauke, Eugene Demler and Gang Chen for helpful discussions. This work is supported by National Key R\&D Program of China grant 2016YFA0301603, NNSFC grant 11874341, the Fundamental Research Funds for the Central Universities, the Anhui Initiative in Quantum Information Technologies, and the Chinese Academy of Sciences.
\end{acknowledgments}

\textbf{Author contributions} B.Y., Z.-S.Y. and J.-W.P. conceived and designed this research. B.Y, H.-N.D., Z.-S.Y. and J.-W.P built the experimental apparatus. H.S, B.Y, H.-Y.W., Z.-Y.Z. and G.-X.S. performed the experiments and analysed the data. All authors contributed to manuscript preparation.

\textbf{Data availability}
Data for figures that support the current study are available at \href{https://dataverse.harvard.edu/dataset.xhtml?persistentId=doi:10.7910/DVN/GPMPMP}{https://doi.org/10.7910/DVN/GPMPMP}

\textbf{Code availability}
Numerical simulations were performed with python code that makes use of the OpenMPS Library available at \href{https://openmps.sourceforge.io/}{https://openmps.sourceforge.io/} and the ALPS project available at \href{http://alps.comp-phys.org/}{http://alps.comp-phys.org/}. Python source codes are available from the corresponding authors on request.

\textbf{Competing interests}
The authors declare no competing interests.


\newpage
\onecolumngrid
\vspace*{0.5cm}
\begin{center}
    \textbf{METHODS AND SUPPLEMENTARY MATERIALS}
\end{center}
\vspace*{0.5cm}
\twocolumngrid
\setcounter{equation}{0}
\setcounter{figure}{0}
\makeatletter
\makeatother
\renewcommand{\theequation}{S\arabic{equation}}
\renewcommand{\figurename}{Extended Data Fig.}

\subsection{Experimental system}

\begin{figure*}[!htb]
\centering
\includegraphics[width=140mm]{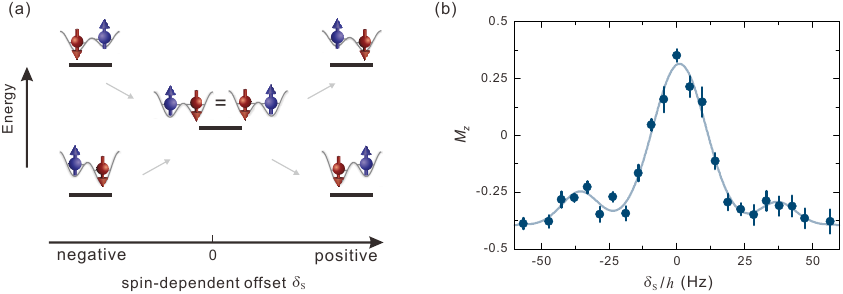}
\caption{Spectroscopy of spin exchange in double wells.
(a) Illustration of the spin-dependent potentials. For two spin configurations in a double well, $\ket{\downarrow,\uparrow}$ and $\ket{\uparrow,\downarrow}$, their energies can be well controlled by tuning the offset value $\delta_s$. For instance, at negative side, the state $\ket{\uparrow,\downarrow}$ has lower energy than $\ket{\downarrow,\uparrow}$.
(b) Spin-exchange under different staggered magnetic potentials.
The states are initialized to $\ket{\uparrow,\downarrow}$ in double wells, corresponding to the staggered magnetization of $M_z=-0.5$. Then, we quench the lattice depth to allow spin exchange dynamics at the coupling strength of $J=26.8(5)$ Hz, holding for 20 ms. The spectroscopy shows a symmetric spin-dependent effect. The solid curve is a guide for the eyes. Error bars are standard deviations throughout this supplementary material.}
\label{FigS1}
\end{figure*}

Our experiment begins with a nearly pure Bose-Einstein condensate of $^{87}$Rb prepared in the hyperfine state of $5S_{1/2}\left|F=1,m_F=-1\right>$.
Then, we increase the confinement along the $z-$axis and load the atoms into a single node of a 3-$\mu m$ spacing blue-detuned lattice to create a quasi-2D quantum gas.
The confinement in the $x-y$ plane is provided by a red-detuned optical dipole trap propagating along the $z$ direction.
To prepare a low-entropy Mott insulator, we implement a newly-developed cooling technique in optical lattices.
In the cooling process, the entropy of the Mott insulator is transferred into the contacting superfluid reservoirs.
In our region of interest (36 copies of 70-site chains), the probability of unity-filling for the Mott insulator is 0.992(1).

To initialize the N\'eel order along the $x$-direction, we employ a site-selective spin addressing technique to overcome the optical diffraction limit.
The atomic Mott insulator is transferred into an spin-dependent superlattice \cite{Yang:2017a} which breaks the odd-even symmetry of the lattice sites and thereby separates the atoms into two subsystems.
For odd and even subsystems, the lattice potentials are controlled to split their transition frequencies between $\ket{\downarrow}$ and $\ket{\uparrow}$ hyperfine levels.
By applying an edited microwave pulse, we flip the atoms of odd sites into $\ket{\uparrow}$ with an efficiency of 0.995(3) through a rapid adiabatic passage.
For each site, the preparation fidelity of the N\'eel state can be estimated as 0.987(3).
Therefore, in the 70-site chain, a defect-free N\'eel order is achieved with a probability of $\sim$ 40\%.

\subsection{Staggered potentials}

The superlattice naturally forms staggered potentials for our ground-band ultracold atoms.
Here, we make use of two distinct effects of the superlattice, which are the spin-dependent and the  spin-independent effects.
Such an optical potential can be expressed as,

\begin{equation}
V(x) = V_{s,\sigma}\cos^2(k_s x+\theta_\sigma) + V^{\text{offset}}_{s,\sigma} - V_l\cos^2(k_s x/2 + \theta),
\end{equation}
where $k_s=2\pi/\lambda_s$ is the wave vector of the short-lattice laser; $\theta_\sigma$ is the spin-dependent phase shift of the short-lattice, and $\sigma$ denotes the two spin states.
The original trap depth of short-lattice $V_s$ is also altered to two spin-dependent terms $V_{s,\sigma}$, and $V^{\text{offset}}_{s,\sigma}$.
At $\theta_{\sigma} = 0$, the superlattice is simplified to the spin-independent form, $V^{\text{offset}}_{s,\sigma} =0$.
By tuning the applying voltage of an electro-optical modulator, we can control the spin-dependent energy with a bandwidth larger than 20 kHz \cite{Yang:2017a}.

The spin-dependent energy shift can be precisely controlled in our system.
To calibrate the zero-energy point $\delta_s =0$, we perform a spectroscopy measurement of spin-exchange dynamics in isolated double-wells.
We prepare the atoms in $\ket{\uparrow,\downarrow}$ state and then quench the superlattice to $V_s=16.8(1) E_r$, $V_l=10.3(1) E_r$ and $\theta=0$.
Here, $E_r = h^2/(2 m \lambda_s^2)$ is the recoil energy with $h$ the Planck constant and $m$ the atomic mass.
In this balanced structure, the resonating superexchange frequency within double wells is $26.8(5)$ Hz.
The inter-well coupling is forbidden by the long-lattice barrier.
The degeneracy of the states $\ket{\uparrow,\downarrow}$ and $\ket{\downarrow,\uparrow}$ is lifted by a nonzero spin-dependent effect.
As we scan the $\delta_s$ and monitor the staggered magnetization $M_z$, a clear spectroscopy of the spin exchange dynamics is observed, as shown in Extended Data Fig. \ref{FigS1}.
The position of the central peak can be determined with a precision of $\sim$0.4 Hz.
In addition, the energy offsets of the side peaks agree well with our calculations based on the lattice structure.

\subsection{Establishing the isotropic Heisenberg Hamiltonian}

We design a special 1D superlattice along the $x$-direction for realizing the isotropic antiferromagnetic Heisenberg model.
When $V_l \ll V_s$, the inter-well coupling becomes comparably strong as the intra-well coupling.
In this 1D chain, there are two types of links connecting the neighboring sites, which can be denoted as odd-even (type I) and even-odd (type II) links.
At $\theta = \pi/4$, these two types of links are identical to each other, while the spin-dependent effect will disappear in this configuration.
To enable the spin-dependent term, we set the superlattice phase to $\theta = \pi/3$ and keep the long-lattice depth at $V_l = 0.44(1) E_r$.
The tunneling strength on the type II links $t_2$ is roughly 10 percent larger than $t_1$ on the type I links \cite{Walters:2013s}.

In the second-order perturbation theory, the spin exchange coupling along these two types of links are,

\begin{equation}
J_{1}=-\frac{4t_{1}^2U}{U^2-(\delta_0-\Delta)^2}, \ \
J_{2}=-\frac{4t_{2}^2U}{U^2-(\delta_0+\Delta)^2}.
\end{equation}
Here, $\delta_0$ and $\Delta$ represent energy offsets between neighboring sites due to the spin-independent term and a linear potential, respectively.
$t_1$ and $t_2$ are tunneling strengths along the two types of nearest-neighbor links.
Since the lattice depths are almost the same for the two spin states, the tunneling strengths are independent of spin components. 
$U$ corresponds to the on-site interaction for all the spin states.
Here we neglect the slight spin-dependence of scattering lengths for $^{87}$Rb and adopt $U_{\uparrow\downarrow}=U_{\uparrow\uparrow}=U_{\downarrow\downarrow}=U$.
When $U < \delta_0 - \Delta \ \text{and} \ \delta_0 + \Delta$, we reverse the sign of $J_1$ and $J_2$ to positive, establishing an antiferromagnetic spin-exchange coupling \cite{Trotzky:2008}.

\begin{figure*}
\centering
\includegraphics[width=150mm]{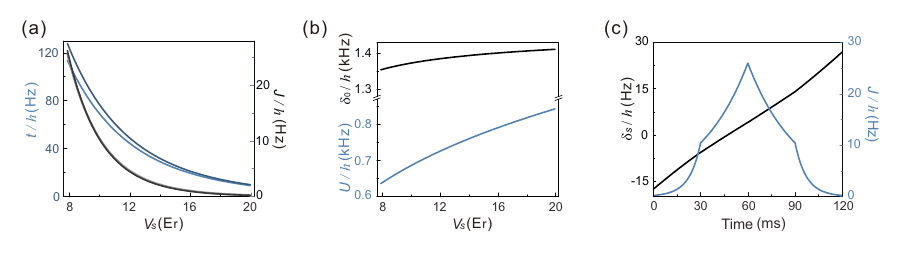}
\caption{ From Bose-Hubbard model to Heisenberg spin model.
(a) The blue and black curves show the tunneling and spin-exchange coupling strengths, respectively.
The lighter colors correspond to $t_1$ and $J_1$ in the type I links, while the darker colors correspond to $t_2$ and $J_2$ in the type II links.
The discrepancy between $J_1$ and $J_2$ has been mostly compensated by the linear potential.
(b) The blue and black curves show the on-site interaction $U$ and the spin-independent staggered potential $\delta_0$ under different short-lattice depths.
(c) The ramping curves of $J$ and $\delta_s$ in the adiabatic passage.}
\label{FigS2}
\end{figure*}

The difference between the spin exchange couplings $J_1$ and $J_2$ is compensated by the state-independent linear potential.
Since our $x$-axis lies at an angle of 4 degree related to the horizontal plane, the gravity leads to a constant 57-Hz/site gradient along the 1D chain.
Consequently, the discrepancy between $J_1$ and $J_2$ is less than 2 percent [see Extended Data Fig. \ref{FigS2}(c)], one order of magnitude lower than the case without this compensation ($J_1$ and $J_2$ would have 20 percent difference in the absence of such a linear potential).

Under our experimental conditions, the Hubbard parameters satisfy such relations $t_1, t_2 \ll \delta_0, U$, as shown in Extended Data Fig. \ref{FigS2}.
We should stress that even though atoms tend to occupy the low-energy sites owing to the non-uniform potential along the 1D chain, the atoms cannot overcome the staggered potential via tunneling in our experimental relevant time scale. Therefore, the Bose-Hubbard Hamiltonian can be mapped into a spin model as,

\begin{equation}
\hat{H}=\sum_{i}(J_{1}\hat{S}_{2i-1}\cdot\hat{S}_{2i} + J_{2}\hat{S}_{2i}\cdot\hat{S}_{2i+1}) + \delta_s\sum_{j}(S_{2j}^z-S_{2j+1}^z).
\end{equation}
By neglecting the slight difference between $J_1$ and $J_2$, we can adopt the geometric mean of them as $J=\sqrt{J_{1}J_{2}}$, and write the Hamiltonian as an isotropic Heisenberg model.

We can understand the spin model in two extreme cases.
In the limit of $\delta_s/J \rightarrow \pm \infty$, the two types of N\'eel orders are the ground states of the system.
In another limit, $\delta_s/J =0$, unlike the Ising-type spin model, this Heisenberg model has a single energetic ground state $\ket{\psi_{\text{AF}}}$.
The SU(2) symmetry of this isotropic Hamiltonian leads to the same rotational symmetry of this ground state.

\subsection{Adiabatic passage}

Starting from the initial N\'eel-ordered state $\left|\psi_0\right>=\left|\uparrow\downarrow\uparrow\downarrow\cdots\right>$, we experimentally optimize the ramping curves to minimize the non-adiabaticity to approach the Heisenberg antiferromagnet $\ket{\psi_{\text{AF}}}$.
The $y-$lattice with 42.0(3) $E_r$ isolates the system into 36 copies of 70-site 1D chains.
The superlattice phases are switched to $\theta = \pi/3$ before ramping down the short-lattice potential.

Then, we slowly ramp the lattice potentials in terms of the Hubbard parameters.
At 0 ms, the staggered magnetic gradient is $\delta_s/h=-17.6(3)$ Hz, and the coupling strength is $J/h= 0.24(1)$ Hz.
During the 60 ms ramping, we separate the curve into two stages with 30-ms intervals, as shown in Extended Data Fig. \ref{FigS2}(c).
The coupling strength and the staggered gradient are increased gradually during the sweep.
At 60 ms, $\delta_s$ is slightly larger than 0, but the zero value of the staggered magnetization $M_z = 0.00(2)$ indicates the achievement of a Heisenberg antiferromagnet.
In another 60 ms, we ramp back the lattice potentials slowly and increase $\delta_s$ oppositely.
The system is driven towards the opposite N\'eel order with $M_z = 0.27(1)$, implying the low-entropy property of the antiferromagnet.
To detect the property of the state at each time, we freeze the state by quenching the lattice barrier to disable the atom dynamics.

\begin{figure}
\centering
\includegraphics[width=76mm]{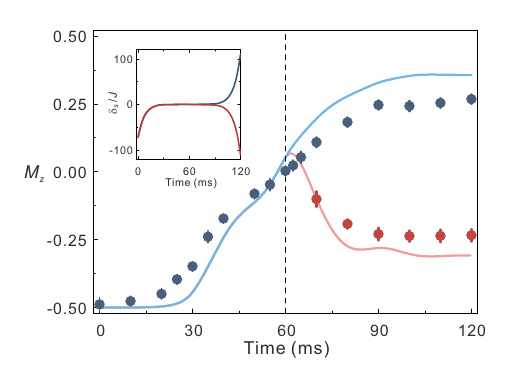}
\caption{Spin dynamics under two types of ramping protocols. Besides the results in Fig. \ref{fig3}a, here the red circles show the results of another sweep process where $\delta_s/J$ is ramped reversely back to a large negative value. The solid curves are the $t$-DMRG simulations of a 70-site spin chain. The inset shows the ramping curves of $\delta_s/J$ in both cases.
}
\label{FigS3}
\end{figure}

Starting from the prepared antiferromagnet, we apply another adiabatic protocol to the system by ramping reversely $\delta_s$ after 60 ms to a large negative value (see Extended Data Fig. \ref{FigS3}). Finally, $M_z$ can return to $-0.24(2)$, which means the decoherence and nonadiabaticity of the system can be well distinguished.

\begin{figure*}
\centering
\includegraphics[width=145mm]{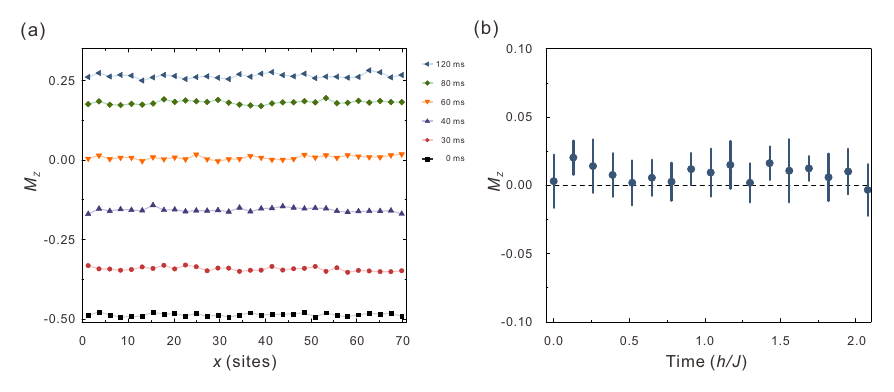}
\caption{Translational invariance and evolution of $M_z$.
(a) Spatial-resolved $M_z$ during the adiabatic passage. For each evolution time, the result is an average of measurement over 3500 copies of 70-site chains.
(b) Evolution of the state in terms of $M_z$. The prepared antiferromagnet is held in a staggered superlattice with $J/h=26.0(5)$ Hz and $\delta_s=0$.
}
\label{FigS4}
\end{figure*}

We quantify the translational symmetry of the antiferromagnet.
Extended Data Fig. \ref{FigS4}(a) shows that the measured $M_z$ does not gain any spatial periodicity.
Such phenomena is consistent with the translational-invariant property of the antiferromagnet.
Besides, we hold the atoms in the strong coupling regime after the 60-ms adiabatic passage, where $J/h=26.0(5)$ Hz and $\delta_s=0$.
As shown in Extended Data Fig. \ref{FigS4}(b), $M_z$ maintains almost constant and does not exhibit any oscillations, which suggests that the created state is close to thermal equilibrium.

\subsection{Nearest-neighbor spin correlation}

The $in\ situ$ atomic densities are detected via an absorption imaging with $\sim$1 $\mu$m spatial resolution.
Even though the imaging process is destructive, we can record the density distributions of both spin states ($\ket{\uparrow} / \ket{\downarrow}$) in two successive images.
We choose a region of interest (ROI) containing 36 copies of 70-site 1D chains in the center of the atomic cloud for the statistical analysis.

The nearest-neighbor correlation function is defined as,

\begin{equation}
C(d=1)= \frac{4}{N}\sum_{j}\left(\left<\hat{S}_j^z\hat{S}_{j+1}^z\right> - \left<\hat{S}_j^z\right>\left<\hat{S}_{j+1}^z\right>\right),
\end{equation}
where $N$ represents the number of sites.
In our spin-balanced system, the second term in the sum is equal to the square of the staggered magnetization, as $-1/N\sum_{j}\langle\hat{S}_j^z\rangle\langle\hat{S}_{j+1}^
z\rangle= M_z^2$.
The first term in the sum is obtained by averaging the correlated measurements on two types of links, as $\frac{1}{2}\left(\langle\hat{S}_{2j-1}^z\hat{S}_{2j}^z\rangle + \langle\hat{S}_{2j}^z\hat{S}_{2j+1}^z\rangle \right)$.
The correlation functions are given as following,

\begin{equation}
\begin{split}
&\text{Type I :} \ \ \\
&-\frac{4}{N}\sum_{j}\langle\hat{S}_{2j-1}^z\hat{S}_{2j}^z\rangle =\frac{1}{2} \left(P_{\ket{\uparrow,\downarrow}}^{\text{I}} + P_{\ket{\downarrow,\uparrow}}^{\text{I}} -P_{\ket{\uparrow,\uparrow}}^{\text{I}}- P_{\ket{\downarrow,\downarrow}}^{\text{I}}\right),\\
\\
&\text{Type II :}  \ \ \\
&-\frac{4}{N}\sum_{j}\langle\hat{S}_{2j}^z\hat{S}_{2j+1}^z\rangle =\frac{1}{2} \left(P_{\ket{\uparrow,\downarrow}}^{\text{II}} + P_{\ket{\downarrow,\uparrow}}^{\text{II}} -P_{\ket{\uparrow,\uparrow}}^{\text{II}}- P_{\ket{\downarrow,\downarrow}}^{\text{II}}\right).
\end{split}
\end{equation}
Here, $P^{\text{I(II)}}$ represent the probabilities of spin states on the corresponding links.
Making use of the superlattice, we can combine the atoms on neighboring sites into a single lattice well.
These two types of links are measured at two kinds of superlattice configurations, one has $\theta=0$ and the other has $\theta=\pi/2$.
Depending on the atom number on the combined site, we can derive the desired probabilities.

\begin{figure*}
\centering
\includegraphics[width=140mm]{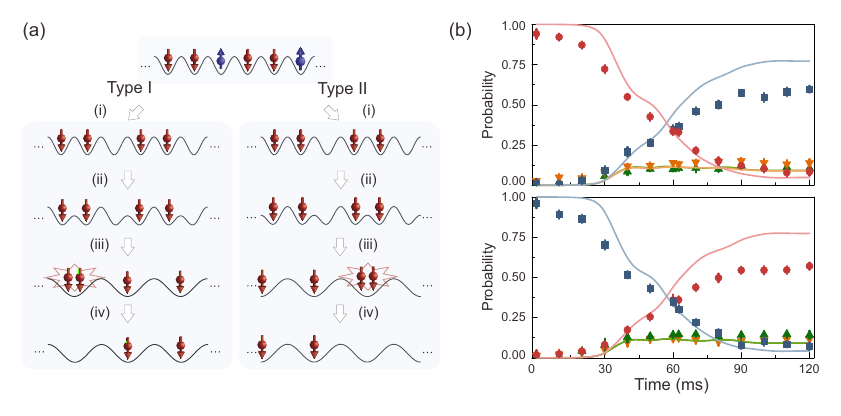}
\caption{Detection of the nearest-neighbor spin correlations.
(a) The detection scheme for the spin components on two types of nearest-neighbor links. In step (i), all $\ket{\uparrow}$ atoms are removed by a resonating imaging light pulse.
Then in (ii), the atoms on type I or type II links are respectively loaded into superlattice double-well units with $\theta=0$  and $\theta=\pi/2$.
In step (iii), the atoms within each double-well are combined and doublons in the long-lattice site get lost after the PA collision.
Finally, in step (iv), we count the remaining atoms and deduce the probabilities of the state $\ket{\downarrow,\downarrow}$.
(b) The upper (lower) graph shows the state probabilities on type I (II) links.
In these two graphs, red circles, blue squares, green triangles and yellow inverted triangles represent the probabilities of $\ket{\uparrow,\downarrow},\ket{\downarrow,\uparrow},\ket{\uparrow,\uparrow}$ and $\ket{\downarrow,\downarrow}$, respectively.
The solid curves are numerical caculations of a 70-site spin chain.}
\label{figS5}
\end{figure*}

We probe the parity of site occupations by applying a photoassociation (PA) laser to the atomic sample.
The laser is red-detuned to the D2 line of $^{87}$Rb by 13.6 $\rm{cm}^{-1}$, which has an intensity of 0.56 $\rm{W/cm}^2$ and lasts for 20 ms.
Assisted by this laser, atom pairs in hyperfine state $\ket{\downarrow}$ can be excited to the $\nu=17$ vibrational state of the $0_g^-$ long-range molecular channel \cite{Fioretti:2001s}.
This unstable molecule would decay back to free atoms and gain kinetic energy to escape from the trap.
Therefore, the ratio of atom loss is equivalent to the probability of the state $\ket{\downarrow,\downarrow}$ before the combining operation.
The detection scheme is illustrated in Extended Data Fig. \ref{figS5}(a), where two sets of measurements are carried out for these two types of links.
In this way, we obtain the probabilities of $P_{\ket{\downarrow,\downarrow}}^{\text{I,II}}$, as shown in Extended Data Fig. \ref{figS5}(b).

For the other spin states $\ket{\uparrow,\uparrow},\ \ket{\uparrow,\downarrow},\ \ket{\downarrow,\uparrow}$, we employ the site-selective spin addressing technique to transfer them into the  $\ket{\downarrow,\downarrow}$ state before implementing the detection procedures \cite{Yang:2017a}.
For instance, we flip all of the atoms on odd sites for probing the probabilities $P_{\ket{\uparrow,\downarrow}}^{\text{I}}$ and $P_{\ket{\downarrow,\uparrow}}^{\text{II}}$.
Finally, the probabilities of these four states allow us to calculate the spin correlation $C(d=1)$.
However, all of the measurement errors would propagate into the final results.
We use the basic normalization condition $P_{\ket{\uparrow}}+P_{\ket{\downarrow}} =1$ to reduce the errors by only taking the residual atoms in the last step of Extended Data Fig. \ref{figS5}(a) into calculations.

In Extended Data Fig. \ref{figS5}(b), the probabilities of the four spin state for type I and type II nearest-neighbor links are depicted.
The deviation from the numerical calculations is mainly caused by the motional excitations in Hubbard model and the lattice heating during the adiabatic passage.
Besides, the inefficiency of the state manipulations accumulates and finally reduces the detected spin correlations.

\subsection{Motional excitations}

\begin{figure*}[!htb]
\centering
\includegraphics[width=70mm]{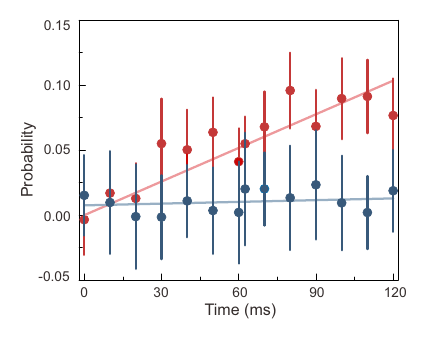}
\caption{Motional excitations during the adiabatic passage. The blue (red) circles represent the ratios of the doublon excitations with identical (opposite) spin components. The solid curves are linear fits to the experimental data.}
\label{figS6}
\end{figure*}

Besides spin excitations, the particle-hole excitations emerge when the atoms tunnel to their neighboring sites.
Such motional excitations are detected by counting the ratio of doublons $\ket{\downarrow\downarrow}$ making use of the parity-projection technique.
Using a global spin-flip, we can acquire the doublons on the state $\ket{\uparrow\uparrow}$ as well.
However, atoms have to tunnel to their next-nearest-neighbor sites in order to create such identical-spin doublons.
Extended Data Fig. \ref{figS6} shows that the excitations to the state $\ket{\uparrow\uparrow}$ and $\ket{\downarrow\downarrow}$ are generally rare [1(1) percent] in the adiabatic passage.

The more probable motional excitations are original from the atoms that tunnel to their neighboring sites and therefore result in the $\ket{\uparrow\downarrow}$ doublons.
From the normalization condition, the ratio of these excitations can be deduced as $1-\sum_{\mu,\nu}P_{\ket{\mu,\nu}}$, with $\mu,\nu=\ \uparrow,\downarrow$.
As shown in Extended Data Fig. \ref{figS6}, the ratio of this type of motional excitations grows gradually during the state evolution, which is 0.04(3) for our target state.
The propagation of such excitations is strongly suppressed by the spin-independent staggered offset $\delta_0$.
Besides, the linear potential also suppresses the long-range motional transport \cite{Yang:2020s}.

\subsection{Next-nearest-neighboring spin correlation}

\begin{figure*}[!htb]
\centering
\includegraphics[width=150mm]{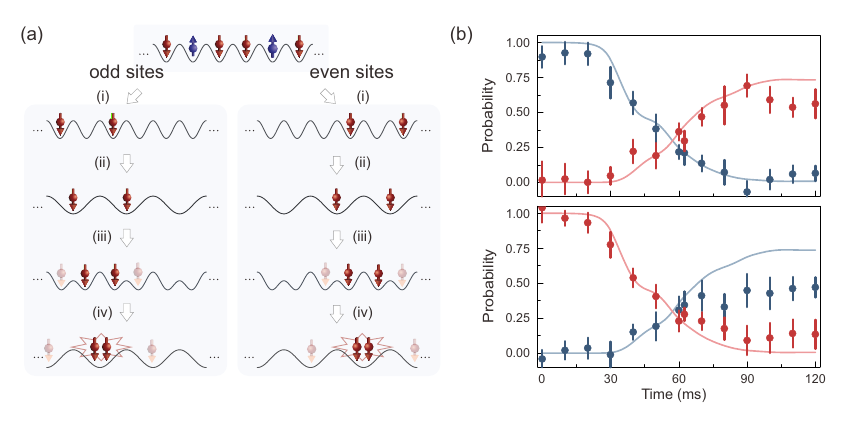}
\caption{Detection of next-nearest-neighbor spin correlations.
(a) The detection scheme for resolving the probabilities of state $\ket{\downarrow;\downarrow}$ on two types of links.
For probing the correlations on odd (even) sites, in step (i), we remove all the $\ket{\uparrow}$ atoms as well as the atoms residing on even (odd) sites.
In step (ii), the remaining atoms are transferred into the long-lattice potential.
Then in step (iii), the atoms are split into adjacent short-lattice sites with equal possibilities.
Finally, in step (iv), we combine the atoms and count the loss of doublons by removing them in the long-lattice potential.
(b) Probabilities of the states $\ket{\uparrow;\uparrow}$ and $\ket{\downarrow;\downarrow}$.
The upper and lower graphs show the evolution of spin components for neighboring odd and even sites, respectively.
The blue circles represent the probabilities of the state $\ket{\uparrow;\uparrow}$, and the red circles correspond to the state $\ket{\downarrow;\downarrow}$.
The solid curves are the numerical calculations of a 70-site spin chain.}
\label{figS7}
\end{figure*}

According to the Mermin-Wagner-Hohenberg theorem \cite{Mermin:1966s}, the 1D isotropic Heisenberg Hamiltonian cannot support a long-range spin order \cite{Yang:2017s}.
As the distance grows, the magnetic correlation decays rapidly in the 1D Heisenberg antiferromagnetic state.
So the occurrence of nonzero next-nearest-neighbor spin correlation would be a distinctive signature of approaching a low-entropy antiferromagnet.

However, without a site-resolved imaging, the superlattice structure normally only provides the access to the nearest-neighbor spin correlation because of its $Z_2$ translational symmetry \cite{Trotzky:2010s,Dai:2016s}.
Here, owing to our precise control of the superlattice phase, we can extend the capability of the correlated spin measurement to next-nearest-neighbor links.

Extended Data Figure \ref{figS7}(a) shows the procedure of detecting the probability of spin state $\ket{\downarrow;\downarrow}$ in next-nearest-neighboring links.
After a selective atom removal, the superlattice serves as an array of atomic splitters to redistribute the remaining atoms into their adjacent sites with equal probabilities.
Subsequently, changing the superlattice phase $\theta$ by $\pi/2$, we can transfer the $\ket{\downarrow}$ atoms originally lying two sites apart to the same long-lattice unit in a probabilistic way.
The atom loss in a PA collision can be used to deduce the probability of spin state $\ket{\downarrow;\downarrow}$.
Similarly, the $\ket{\uparrow; \uparrow}$ state on next-nearest-neighbor sites can be detected with an additional global spin-flip pulse before the step (i).

We define a next-nearest-neighbor correlator as,

\begin{equation}
C'(d=2)=P_{\ket{\uparrow;\uparrow}}+P_{\ket{\downarrow;\downarrow}},
\end{equation}
where $P_{\ket{\uparrow;\uparrow}}$ and $P_{\ket{\downarrow;\downarrow}}$ represent the probabilities of spin state $\left|\uparrow;\uparrow\right>$ and $\left|\downarrow;\downarrow\right>$ on next-nearest-neighbor links.
These probabilities equal to four times of the corresponding atom losses in the PA collision.
For determining the probabilities accurately, we subtract the contribution of the motional excitations in the PA collision.

Extended Data Figure \ref{figS7}(b) shows the variations of the probabilities $P_{\ket{\uparrow;\uparrow}}$ and $P_{\ket{\downarrow;\downarrow}}$ for odd- and even-sites respectively.
Since only a quarter of atoms contribute to the signal, the probabilities measured here have larger errors compared to their counterparts on nearest-neighbor links.

\subsection{Numerical simulation}

\begin{figure*}[!htb]
\centering
\includegraphics[width=140mm]{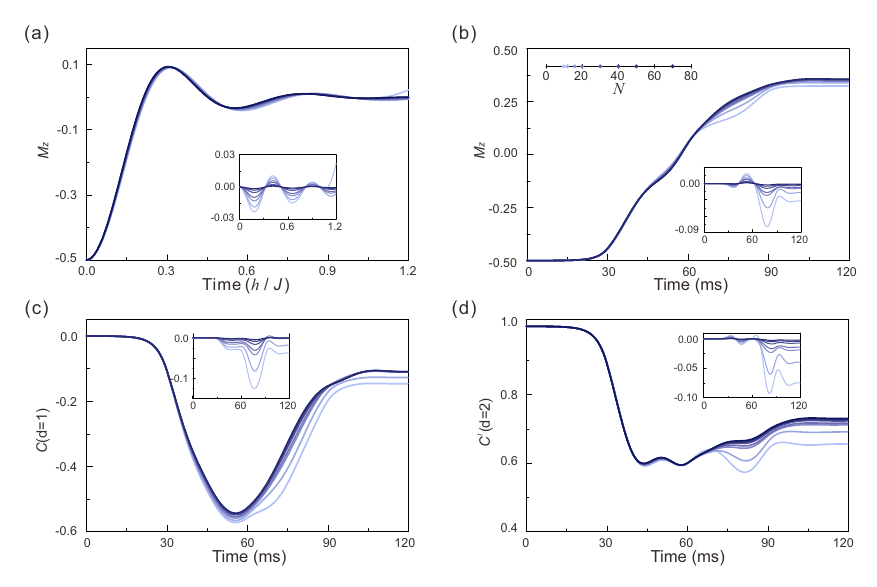}
\caption{Numerical calculations of the Heisenberg spin chains. (a) spin relaxation dynamics after quenching the N\'eel order to the Heisenberg spin model.
(b)(c) and (d) are respectively the numerical results of the evolution of $M_z$, $C(d=1)$ and $C'(d=2)$ in the adiabatic passage.
Different blue colors represent the chain length of 10, 12, 16, 20, 30, 40, 50 and 70, respectively.
The insets show the discrepancies between the numerical results with the system at $N$=70.
The finite-size convergence is well below our detection errors.}
\label{figS8}
\end{figure*}

\begin{figure*}[!htb]
\centering
\includegraphics[width=116mm]{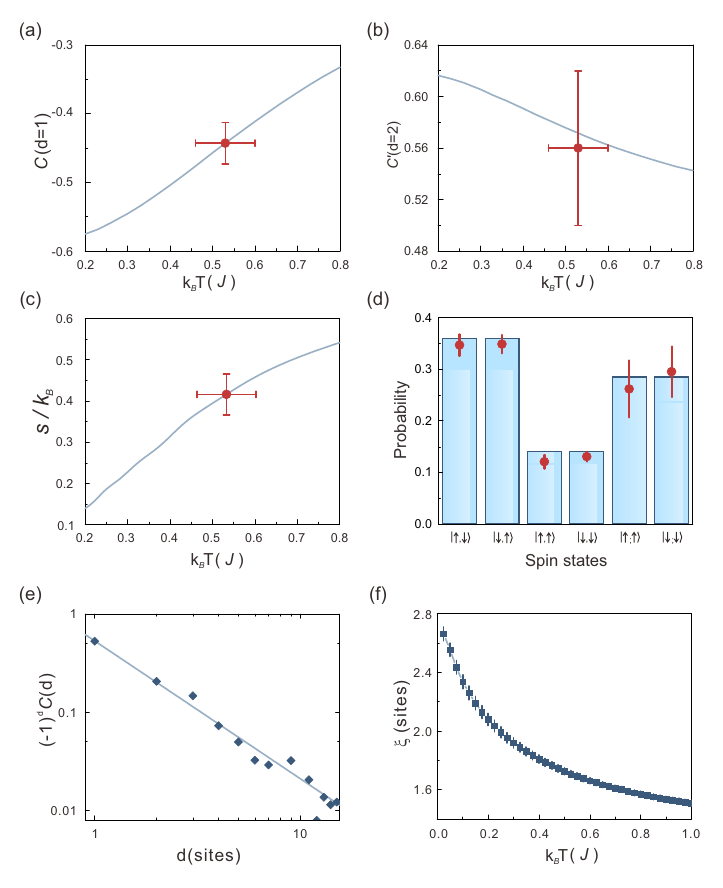}
\caption{Numerical simulations. We theoretically simulate the 70-site spin chain with the finite-temperature QMC methods and the $t$-DMRG methods.
(a), (b) and (c) show the spin correlations $C(d=1)$,  $C'(d=2)$ and the entropy per particle versus temperatures, respectively.
The red circles in the plots represent the prepared target state.
(d) Probabilities of the spin states in the nearest- and next-nearest-neighbor links.
The data points denote the measured probabilities of spin components, and the blue bars are the theoretical predictions of a thermalized Heisenberg antiferromagnet at $k_B T = 0.53J$.
(e) Correlation function. For the state prepared after a 60 ms ramp, we calculate the correlation functions $(-1)^d C(d)$ using the $t$-DMRG method. The solid curve represents a fitting of the calculations with a power-law function.
(f) Based on the power-decay behavior in (e), we obtain the correlation length $\xi$ [spin correlation decays to $C(d=1)/e$] of the system under different temperatures. The correlation length does not diverge even when the temperature approaches zero.
}
\label{figS9}
\end{figure*}

Based on the Heisenberg spin model, the numerical results for both the post-quench dynamics and the adiabatic passage are calculated using the $t$-DMRG algorithm in the framework of matrix product states.
In the numerics, we neglect the slight difference between the spin-exchange couplings of the two types of links and adopt their geometric mean $J=\sqrt{J_{1}J_{2}}$ in the spin model.
We implement the methods using the OpenMPS library \cite{Jaschke:2018s} and fix the maximal bond dimension to $D=1500$.
Convergence are achieved at a time step of $10^{-2}\ h/J$ and $3\times10^{-4}$ s for the quench dynamics and the adiabatic passage, respectively.
The truncation threshold is $10^{-6}$ per time step.
We investigate the finite-size effects by calculating the spin dynamics in 1D systems at several sizes $N$.
As shown in Extended Data Fig. \ref{figS8}, the discrepancy between the curves for $N = 40$ and $N = 70$ is on the order of $10^{-2}$, indicating the volume convergence of our calculations.

Temperature $T$ and entropy $\mathcal{S}/N$ of the state are estimated by comparing the experimental results with finite-temperature quantum Monte Carlo methods (QMC) \cite{Bauer:2011s}.
All the simulations are carried out in the grand-canonical ensemble on a 70-site open chain.
In such a large system, the finite-size effect can be eliminated.
After $10^5$ times of thermalization, we continue to perform the QMC sweeps for at least $5\times10^6$ times to get the results under each setting.
We choose the temperature $k_B T/J$ from 0.2 to 0.8 with a 0.03 interval.

From the nearest-neighbor correlation of interest [$C(d=1)=-0.44(3)$], we derive the temperature of our system to be 0.53(7)$J$, see Extended Data Fig. \ref{figS9}(a).
In Extended Data Fig. \ref{figS9}(b), we present the next-nearest-neighbor correlation as a function of the temperature.
At a temperature of 0.53(7)$J$, $C'(d=2)$ reaches 0.57(1) in this thermal state.
The entropy per particle is obtained using another QMC method, which is called quantum Wang-Landau algorithm \cite{Wang:2001s}.
The cutoff of series expansion is set to be $10^3$.
As shown in Extended Data Fig. \ref{figS9}(c), the entropy per particle deduced from the temperature of the spin chain is 0.42(5)$k_B$.

Based on the comparison between the measured probabilities of the spin state and the prediction of the thermalized state, our system is inferred to be a thermalized state, see Extended Data Fig. \ref{figS9}(d).

From the $t$-DMRG simulations, we plot the spin correlations of the antiferromagnet up to 15 sites.
As shown in Extended Data Fig. \ref{figS9}(e), the correlation function exhibits a power-law decay as the distance $d$ increases, which is consistent with the property of the Tomanaga-Luttinger Liquid \cite{Haldane:1981,Giamarchi:2004,Yang:2017s}.

\subsection{Fluctuations of the staggered magnetization}

The spin fluctuations in terms of $\delta M_z$ can reflect non-local spin correlations of many-body states.
However, the detection of $\delta M_z$ is obstructed by the detection noise, for instance the shot noise of our absorption imaging.
Here, we develop a method to resolve the spin fluctuations on the basis of the two-image measurements.
The total density $n_A+n_B$ corresponds to a unit-filling Mott insulator with vanishing density fluctuations.
Since the staggered magnetization is $M_z=n_A/(n_A+n_B)-1/2$, where the spin fluctuations only exist on $n_A$.
In this respect, we can extract $\delta M_z$ directly from the fluctuations of $n_A$.
For reference, we consider the background fluctuations due to the photon shot-noise by analyzing the total density $n_A+n_B$.
We resolve the spin fluctuations $\delta M_z$ by subtracting the shot-noise background, as $\delta M_z\propto\sqrt{\sigma(n_A)^2-\sigma_{bg}^2}$.

\begin{figure*}[!htb]
\centering
\includegraphics[width=140mm]{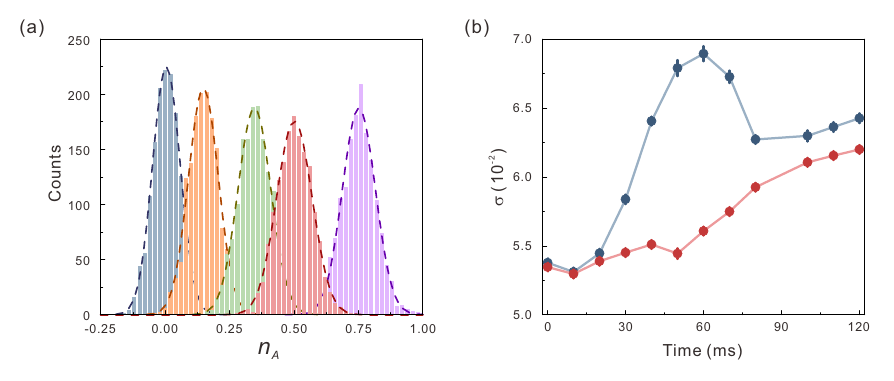}
\caption{Spin fluctuations.
(a) Histograms of $n_A$. We count the density among 2.37 copies of 70-site chains (one pixel along $y$-axis corresponds to 2.37 lattice sites).
The blue, orange, green, red and purple histograms correspond to the state at the evolution time of 0 ms, 30 ms, 40 ms, 60 ms and 120 ms in the adiabatic passage, respectively.
For each measurement, 1500 samples are taken for the statistical analysis. The dashed curves are fittings based on the Gaussian function.
(b) Density fluctuations (blue) and background noises (red) throughout the adiabatic passage.
}
\label{FigS10}
\end{figure*}

As Extended Data Fig. \ref{FigS10} (a) shows, the density histogram of $n_A$ throughout the adiabatic sweep are obtained by counting the densities on 2.37 copies 70-site chains (one pixel equals to 2.37 sites).
Based on 100 experimental repetitions, we acquire 1500 groups of measurement for each histogram.
Here, $\sigma(n_A)$ represents the Gaussian waist of the histogram of $n_A$ (the Gaussian function is used for fitting), as shown in Extended Data Fig. \ref{FigS10} (b).
Taking into account the imaging resolution and the pixel size, the fluctuation of staggered magnetization $\delta M_z$ can be extracted.

\end{document}